\def\rfr#1{eq. (\ref{#1})}

\def\asec{$''$ cy$^{-1}$}

\def\dert#1#2{\frac{{{d}}{#1}}{{{d}}{#2}}}              % derivate parziali e totali prima e seconda

\def\asec{$''$ cy$^{-1}$}

\def\ctz#1{Ref.~\cite{#1}}

\def\asec{$''$ cy$^{-1}$}

\def\dert#1#2{\frac{{{d}}{#1}}{{{d}}{#2}}}              % derivate parziali e totali prima e seconda

\def\asec{$''$ cy$^{-1}$}

\def\bar{\begin{eqnarray}}
\def\ear{\end{eqnarray}}
\def\bb{\bibitem}
\def\eqi{\begin{equation}}
\def\eqf{\end{equation}}
\def\eqia{\begin{eqnarray}}
\def\eqfa{\end{eqnarray}}
\def\rp#1#2{\frac{#1}{#2}}
\def\ct#1{\cite{#1}}
\def\lb#1{\label{#1}}

\def\media#1{\left\langle #1\right\rangle}

\def\dotc{\frac{\dot c}{c}}
\def\pare#1{\left(#1\right)}

\def\bds#1{\boldsymbol{#1}}

\documentclass[11pt]{article}%{ws-ijmpd}
\usepackage{amsmath,amsthm,amscd,amssymb}
\usepackage{latexsym}
\usepackage{graphicx,epsfig}

\begin{document}

\noindent{\bf \LARGE{Solar System planetary tests of $\dot c/c$ }}
\\
\\
\\
{L. Iorio}\\
{\it INFN-Sezione di Pisa. Address for correspondence: Viale Unit$\grave{a}$ di Italia 68\\
70125 Bari, Italy
\\tel./fax 0039 080 5443144
\\e-mail: lorenzo.iorio@libero.it}

\vspace{4mm}

%\title{SOLAR SYSTEM PLANETARY TESTS OF  $\dot c/c$}

%\author{LORENZO IORIO}

%\address{INFN- Sezione di Pisa. Viale Unit$\grave{a}$ di Italia 68, 70125 Bari (BA), Italy
%\\e-mail: lorenzo.iorio@libero.it}

% \maketitle

%\begin{history}
%\received{16 June 2008}
%\revised{Day Month Year}
%\accepted{21 September 2008}
%\comby{Jorge Pullin}
%\end{history}

\begin{abstract}
Analytical and numerical calculations show that a putative temporal variation of the speed of light $c$, with the meaning of space-time structure constant $c_{\rm ST}$, assumed to be linear over timescales of about one century, would induce a secular precession of the longitude of the pericenter $\varpi$ of a test particle orbiting a spherically symmetric body. By comparing such a predicted effect to the corrections $\Delta\dot\varpi$ to the usual Newtonian/Einsteinian perihelion precessions of the inner planets of the Solar System,  recently estimated by E.V. Pitjeva by fitting about one century of modern astronomical observations with  the standard dynamical force models of the EPM epehemerides, we obtained $\dot c/c =(0.5\pm 2)\times 10^{-7}$ yr$^{-1}$. Moreover, the possibility that $\dot c/c\neq 0$ over the last century is ruled out at $3-12\sigma$ level by taking the ratios of the perihelia for different pairs of planets. Our results are independent of any measurement of the variations of other fundamental constants which may be explained by a variation of $c$ itself (with the meaning of electromagnetic constant $c_{\rm EM}$).
%They retain their validity also for a variation of the Newtonian gravitational constant $G$ provided that $-2\dot c/c$ is replaced by $\dot G/G$ %throughout the paper.
It will be important to repeat such tests if and when other teams of astronomers will estimate their own corrections to the standard Newtonian/Einsteinian planetary perihelion precessions.
\end{abstract}


\section{Introduction}
In this paper we will deal with the problem of effectively putting on the test an hypothetical time-variation of the speed of light $c$ in a purely phenomenological, model-independent way with local, Solar-System-scale astronomical observations.

Varying Speed of Light (VSL) theories were proposed in recent times to accommodate certain features of the hot Big-Bang cosmology. In this sense, the first modern VSL theory was put forth by Moffat in \ctz{Mof}; for other pioneering works see, e.g., \ctz{Bar} and \ctz{mag}. Since then, this subject was dealt with by many authors investigating different aspects of it; see, e.g., \ctz{Mag} for an extensive review. Broadly speaking, such theories can be subdivided in two categories: those encompassing space-time variations of $c$, motivated by cosmology, and those where $c$ varies with the energy scale, related to phenomenological quantum gravity. Subtle issues concerning fundamental aspects of VSL theories have been recently discussed in \ctz{Uz}, \ctz{El} and \ctz{reply}.

From the observational point of view,  the measured percent change \ct{Webb1,Mur,Webb} \eqi\rp{\Delta\alpha}{\alpha}= (-7.2\pm 1.8)\times 10^{-6}\lb{alf}\eqf
 of the fine structure constant\footnote{For a review on the issue of the variation of $\alpha$ and other fundamental constants, see, e.g., \ctz{Uzan}.}  \eqi \alpha = \rp{q_e^2}{\hslash c},\eqf where $q_e$ is the rationalized electron charge and $\hslash$ is the Planck's constant, respectively, from quasar observations at redshift $z\approx 0.5-3.5$ was very important for VSL theories; indeed, the natural question arises: if $\alpha$ is varying, is such a change due to $q_e$, $h$ or $c$?
 By attributing $\Delta\alpha/\alpha$ to a temporal variation of $c$, it follows
 \eqi\rp{\Delta c}{c}=-\rp{\Delta\alpha}{\alpha}.\eqf
 According to the distinction of the many facets of $c$ proposed in \ctz{Uz}, the $c$ present here would be $c_{\rm EM}$, i.e. the electromagnetic constant.
 Since \eqi\rp{\Delta\alpha}{\alpha}\equiv\rp{ \alpha_{\rm past} -\alpha_{\rm today}}{\alpha_{\rm today}}< 0, \eqf the value of $\alpha$ was lower in the past; thus, $c$ would have been larger in the past and it would be decreasing.
 By assuming a linear time dependence
 \eqi \rp{\Delta c}{c} \approx  \dotc (t-t_0)<0,\lb{dotc}\eqf
 from \rfr{alf}
 it can be obtained
 \eqi\dotc \approx (-8\pm 2)\times 10^{-16}\ {\rm yr}^{-1}\lb{cdot}\eqf
 for
 \eqi t-t_0\approx 9\ {\rm Gyr}\eqf which approximately corresponds to the temporal interval spanned by the data analyzed in \ctz{Webb}, i.e. from 23$\%$ to 87$\%$ of the age of the
universe.
A tighter bound could be obtained from the constrain in the variation of $\alpha$ over the last 1.8 Gyr
\eqi\left|\rp{\dot\alpha}{\alpha}\right|\leq 3\times 10^{-17}\ {\rm yr}^{-1}\lb{oklo}\eqf
from an analysis of the Oklo mine data \ct{Pet}.

Can local\footnote{For a strategy to combine local and cosmological tests of varying fundamentals constants like $\alpha$ and $G$ see \ctz{Sha}.} (in space and time) astronomical observations tell us something about the hypothesis that $c$ undergoes temporal variations?
%
%the measured variation of $\alpha$ is due to $c$?
The answer is, in principle, positive because the motion of the major bodies of the Solar System is governed by the dynamical equations of motion of classical general relativity in which $c$, playing the role of the space-time structure constant  $c_{\rm ST}$ \ct{Uz}, is explicitly present;
 %contrary to $\hslash$ and $q_e$;
 as we will see, a  (slowly) time-varying $c$ induces dynamical effects that can be tested with the latest planetary observations independently of $\alpha$. Of course, it must be borne in mind that such tests can only constrain $\dot c/c$ over timescales of about one century, corresponding to the temporal interval covered by the modern astronomical observations of the major bodies of the Solar System which are used to construct the present-day highly accurate ephemerides.

\section{The dynamical effects of $\dot c/c$ on the orbital motion of a test particle}
We will follow a phenomenological approximation, without working in any specific VSL theoretical framework.
 By inserting \rfr{dotc} into the 1PN gravitoelectric acceleration \ct{Sof} of order $\mathcal{O}(c^{-2})$
  \eqi \bds A_{\rm 1PN} = \rp{GM}{c^2r^3}\left[\left(\rp{4GM}{r} -v^2\right)\bds r + 4(\bds r\bds\cdot\bds v)\bds v\right],\lb{soffi}\eqf
  which causes the well-known Mercury's perihelion precession of 43.98 arcsec cy$^{-1}$,
  one gets
 \eqi \Delta \bds A_{\rm 1PN} \approx \left[-2\pare\dotc(t-t_0)\right]\bds A_{\rm 1PN};\lb{picco}\eqf
 here and in the following $c = c_0=c(t_0)$. Note that, according to the distinction of the many facets of $c$ proposed in \ctz{Uz}, the quantity varying here is the space-time structure constant  $c_{\rm ST}$, which is, in principle, not related to the electromagnetic constant $c_{\rm EM}$. It can be shown that the first term of \rfr{soffi} is proportional to $c_{\rm ST}^2/c^4_{\rm E}$, while the other two terms are proportional to $1/c^2_{\rm E}$, where $c_{\rm E}$ the Einstein space-time matter constant \ct{Uz}; however, in order to get the correct Newtonian limit for gravity \ct{Uz}, we will assume $c_{\rm ST} = c_{\rm E}$.
  Our approach, which has the merit of making direct and unambiguous contact with the observations giving definite answers\footnote{It should be recalled that the observational basis of VSL phenomenology is, at present, quite meager.}, might be criticized from a  theoretical point of view as, perhaps, too na\"{\i}ve; indeed, as pointed out by Jordan \ct{Jor,Jor2}, in general, it is not consistent to allow a constant to vary in an equation that has been derived from a variational principle under the hypothesis that this quantity is constant; one needs to go back to the Lagrangian and derive new equations with the constant treated as a dynamical field.
  However, whatever the temporal evolution of $c(t)$ may be,  the approximation of \rfr{dotc} is adequate for the practical purpose of testing it over relatively short timescales like the last century in which modern astronomical planetary observations were collected. Incidentally, let us note that, in this case, certain observational issues \ct{Uz,El} can be neglected, at least from a practical point of view. Indeed, from \eqi d\tau =\rp{\sqrt{g_{00}}}{c}dt,\eqf where $c$ is the space-time structure constant $c_{\rm ST}$, by assuming a linear time variation of it, the following shift in the measured proper time would occur for a static field\eqi \left|\rp{\Delta \tau}{\tau}\right|=\rp{\dot c}{c}\rp{\Delta t}{2};\eqf over $\Delta t = 100$ yr and by assuming for $c_{\rm ST}$ the same rate of change of $c_{\rm EM}$ obtained from the Oklo natural reactor data for $\alpha$ of \rfr{oklo}, it turns out
 \eqi \left|\rp{\Delta\tau}{\tau}\right|\approx 1.5\times 10^{-15},\eqf which is basically undetectable given the present-day accuracy in realizing the SI second by the Bureau International des Poids et Mesures (BIPM), i.e. \ct{BIPM} $3\times 10^{-15}$.

  From a dynamical point of view, $\Delta\bds A_{\rm 1PN}$ can certainly be considered as a small perturbation with respect to  the Newtonian monopole over timescales of about 100 yr, as it will be a posteriori confirmed by the bound on $|\dot c/c|$ that we will obtain with such a hypothesis; the same holds if one uses \rfr{cdot} derived from $\Delta\alpha/\alpha$. Thus, \rfr{picco} can be treated perturbatively with the standard Gauss \ct{Ber} approach which is valid for any perturbing acceleration, whatever its physical origin may be.
In order to evaluate the orbital effects of a generic small disturbing acceleration $\bds W$, it
 is customarily projected onto an orthonormal frame $K$ co-moving with the test particle. The mutually orthogonal unit vectors $\bds{\hat{ r}}$, $\bds{\hat{ \tau}}$, $\bds{\hat{ \nu}}$ of $K$ pick out the radial, transverse  and normal directions, respectively; $\bds{\hat{r}}$ and $\bds{\hat{\tau}}$ are in-plane with $\bds{\hat{r}}$ directed along the particle's radius vector, while  $\bds{\hat{\nu}}$ is out-of-plane, directed along the orbital angular momentum.
The Gauss equations for the variations of the Keplerian orbital elements are \ct{Ber}
\begin{eqnarray}\lb{Gauss}
\dert a t & = & \rp{2}{n\sqrt{1-e^2}} \left[e W_r\sin f +W_{\tau}\left(\rp{p}{r}\right)\right],\lb{gaus_a}\\
\dert e t  & = & \rp{\sqrt{1-e^2}}{na}\left\{W_r\sin f + W_{\tau}\left[\cos f + \rp{1}{e}\left(1 - \rp{r}{a}\right)\right]\right\},\lb{gaus_e}\\
\dert I t & = & \rp{1}{na\sqrt{1-e^2}}W_{\nu}\left(\rp{r}{a}\right)\cos u,\lb{gaus_i}\\
\dert\Omega t & = & \rp{1}{na\sin I\sqrt{1-e^2}}W_{\nu}\left(\rp{r}{a}\right)\sin u,\lb{gaus_O}\\
\dert\omega t & = &\rp{\sqrt{1-e^2}}{nae}\left[-W_r\cos f + W_{\tau}\left(1+\rp{r}{p}\right)\sin f\right]-\cos I\dert\Omega t,\lb{gaus_o}\\
\dert {\mathcal{M}} t & = & n - \rp{2}{na} W_r\left(\rp{r}{a}\right) -\sqrt{1-e^2}\left(\dert\omega t + \cos I \dert\Omega t\right),\lb{gaus_M}
\end{eqnarray}
where $a$, $e$, $I$, $\Omega$, $\omega$ and ${\mathcal{M}}$ are the semi-major axis, the eccentricity, the inclination, the longitude of the ascending node, the argument of pericentre and the mean anomaly of the orbit of the test particle, respectively. The angle $f$ is the true anomaly reckoning the instantaneous position of the test particle along its orbit with respect to the pericentre, $u=\omega+f$ is the argument of latitude, $p=a(1-e^2)$ is the semi-latus rectum and $n=\sqrt{GM/a^3}$ is the un-perturbed Keplerian mean motion related to the un-perturbed Keplerian orbital period by $P_{\rm b}=2\pi/n$.
 For the following calculations it is more convenient to use the eccentric anomaly\footnote{It is defined by $\mathcal{M}=E-e\sin E$.} $E$ in terms of which the un-perturbed  Keplerian ellipse at epoch $t_0$ can be written as
\eqi r = a(1-e\cos E),\lb{kep_r}\eqf
\eqi \cos f = \rp{\cos E - e}{1-e\cos E},\lb{kep_cosf}\eqf
\eqi \sin f = \rp{\sqrt{1-e^2}\sin E}{1-e\cos E},\lb{kep_sinf}\eqf
To obtain the secular, i.e. averaged over one orbital revolution, effects of $\bds W$, it has to be evaluated onto the unperturbed Keplerian ellipse
with the aid of \rfr{kep_r}-\rfr{kep_sinf} and inserted into the right-hand-side of \rfr{gaus_a}-\rfr{gaus_M}; then, an integration with respect to $t$ over an orbital period has to be performed by using
\eqi \rp{dt}{P_{\rm b}} = \left(\rp{1-e\cos E}{2\pi}\right)dE.\lb{kep_dt}\eqf

\begin{figure}
 \centerline{\psfig{file=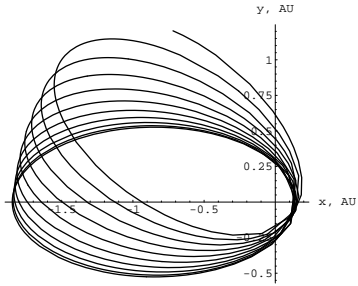,width=4.7cm}}
\vspace*{8pt}
\caption{\footnotesize{Numerically integrated trajectory of a fictitious planet around the Sun affected by the perturbing acceleration of \rfr{picco} in addition to the Newtonian monopole $-GM/r^2$. A positive value large enough ($\dot c/c = 10^4$ yr$^{-1}$) to sufficiently enhance the pericentre precession has been chosen for $\dot c/c$. The initial conditions chosen are   $x_0 = r_{\rm min}=a(1-e)$, $y_0=0$, $z_0=0$, $\dot x_0=0$, $\dot y_0=v_{\rm max}=na\sqrt{(1+e)/(1-e)}$, $\dot z_0=0$ with $a=1$ AU, $e=0.85$: the motion of the planet along the orbit is anticlockwise. The temporal interval spanned by the integration is 10 yr. The retrograde (i.e. clockwise) precession, as predicted by \rfr{dc_dot_o}, is clearly visible.}\label{figura}}
\end{figure}

In the case of \rfr{picco}, with
\eqi t-t_0 = \rp{E-e\sin E}{n},\lb{kep_t}\eqf
the Gauss equation for $\omega$ yields
%
%\eqi \media{\dot a} = -4\pare\dotc\pare{\rp{GM}{c^2}}\left\{ \rp{5}{\sqrt{1-e^2}} -\rp{\left[5 - 3e(1-e)\right]}{1-e^2}  \right\},\lb{dc_dot_a} \eqf
%
%\eqi \media{\dot e} = -2\pare\dotc\pare{\rp{GM}{c^2}}\rp{(1-e^2)}{ae}\left\{ \rp{5}{\sqrt{1-e^2}} -\rp{\left[5 - 7e(1-e)\right]}{1-e^2}  %\right\},\lb{dc_dot_e}\eqf
%
%\eqi  \media{\dot I} = 0, \lb{dc_dot_i}\eqf
%
%\eqi  \media{\dot \Omega} = 0, \lb{dc_dot_O}\eqf
%
\eqi \media{\dot \omega} = -\pare\dotc \left(\rp{GM}{c^2 a}\right)\rp{F(e)}{2\pi},\lb{dc_dot_o}\eqf
%
%\eqi  \media{\dot {\mathcal{M}}} = .\lb{dc_dot_M}\eqf
%
with
\eqi F(e) = \rp{\sqrt{1-e^2}}{e}\int_0^{2\pi}\rp{ 2\left(3+e^2\right)\cos E + e\left(-15 + 7\cos 2E\right)\left(-E + e\sin E\right)   }{\left(1-e\cos E\right)^3}dE\lb{mega};\eqf in Table \ref{tavo} we quote the numerically computed values of $F(e)$ for the inner planets of the Solar System.
\begin{table}
\caption{ \footnotesize{First row: eccentricities $e$ of the inner planets of the Solar System. Second row: numerically calculated values of $F(e)$ according to \protect\rfr{mega}.}}
\centering
\bigskip
\begin{tabular}{ccccc}
\hline\noalign{\smallskip}
%

%{\begin{tabular}{@{}ccccc@{}} \hrule
%
& Mercury & Venus & Earth & Mars\\
\noalign{\smallskip}\hline\noalign{\smallskip}
 $e$ & 0.20563069 & 0.00677323 & 0.01671022 & 0.09341233\\
$F(e)$ & 527.063 & 5862.73 & 2554.07 & 714.504\\
\noalign{\smallskip}\hline
\end{tabular}\label{tavo}
\end{table} %
Note that \rfr{dc_dot_o} holds also for the longitude of pericentre $\varpi=\Omega + \omega$; indeed, since $W_{\nu}=0$, from \rfr{gaus_i} and \rfr{gaus_O} turns out that $\media{\dot I}=\media{\dot \Omega}=0$.
Moreover, it is not possible to attribute the pericentre precession to a re-scaled time-varying gravitational constant because, in this case, also the
Newtonian monopole $-GM/r^2$ would be fictitiously affected.
Our analytical result for the pericentre precession of \rfr{dc_dot_o} is also qualitatively confirmed by a numerical integration of the equations of motion of a fictitious planet around the Sun in which the magnitude of $\dot c/c$ has been purposely set to value large enough to visually inspect the rotation of the orbit in its plane, as depicted in Figure \ref{figura}

Let us stress that \rfr{dc_dot_o} is  different from the precession obtained by Magueijo  by investigating in \ctz{perihel} spherically symmetric solutions to a definite covariant and Lorentz-invariant VSL theory \ct{strunzatill}; indeed, the Magueijo's effect is equal to the usual 1PN precession\footnote{The precession per orbit is shown in \ctz{perihel}; in order to compare it with our results, it must simply be divided by the Keplerian orbital period $P_{\rm b}=2\pi/n$.}
multiplied by an adimensional factor, i.e.
\eqi\dot\omega_{\rm VSL}=-\rp{3nGM}{c^2 a(1-e^2)}\left(\rp{4}{3}\rp{b^2}{\kappa}\right)=-\rp{3(GM)^{3/2}}{c^2 a^{5/2}(1-e^2)}\left(\rp{4}{3}\rp{b^2}{\kappa}\right),\lb{Mague}\eqf where $b$ and $\kappa$ are, in turn, numbers \ct{strunzatill,perihel}, presumably of some fundamental nature, accounting for the dynamical evolution of $c$. Simple dimensional considerations show, in fact, that \rfr{Mague} does not look like the formula one would reasonably expect for a weak-field dynamical precessional effect induced by a (slow) time variation of $c$. Indeed, the basic ingredients that should intuitively enter such a formula  are the two lengths $GM/c^2$ and $a$, characterizing the problem at hand, a possible adimensional function of the eccentricity $e$ and a quantity $Q$ having the dimensions of the reciprocal of time; $[Q] =T^{-1}$. Now, possible candidates for $Q$ are the orbital frequency $n$ and, of course, $\dot c/c$ which is the cause of the effect looked for; excluding quadratic terms in $\dot c/c$, the most natural choice seems to be just $Q=\dot c/c$. Stated differently, it would be possible to express \rfr{dc_dot_o} as the standard 1PN precession times an adimensional factor $\xi$, but the latter one would be
\eqi \xi = -\pare\dotc\rp{1}{6\pi n}(1-e^2)F(e),\eqf where there is only one dimensional parameter related to the variation of $c$, i.e. its percent first derivative, while the other dimensional quantity, specific to the system considered, is the planet's orbital frequency.  As we will see later, the dependence on $a$ and $e$ is crucial for the confrontation with observation-related quantities.

\section{The confrontation with the observations in the Solar System}
By suitably using the perihelia of the inner planets of the Solar System it is possible to  constrain $\dot c/c$ over timescales of the order of about 1 century and even  rule out the  hypothesis that it may have a non-zero value, at least in the last century.

The astronomer E.V. Pitjeva has recently fitted almost one century of planetary data of various types with the dynamical force models of the EPM ephemerides estimating, in the least-square sense, several parameters; the  modelled dynamical features include \ct{Pit05} all the most relevant Newtonian effects (N-body mutual perturbations among the major bodies of the Solar System, Sun's oblateness, 301 large asteroids, massive ring lying in the ecliptic plane accounting for the small asteroids) and the general relativistic Schwarzschild-like accelerations in the harmonic gauge.
Among the various solutions obtained, in one of them she also phenomenologically estimated corrections \ct{Pit05,PitMAG} $\Delta\dot\varpi$ to the standard Newtonian-Einsteinian precessions of the longitudes of the perihelia\footnote{Strictly speaking, the perihelia are not observables; they can be computed from the measured quantities which are ranges, range-rates and angles like right ascension and declination.} of the inner planets by keeping the usual PPN parameters fixed to their general relativistic values. By construction, such corrections $\Delta\dot\varpi$, shown in Table \ref{tavo2}, account, in principle, for any  standard, i.e. Newtonian and/or general relativistic, or exotic un-modelled/mis-modelled forces.
\begin{table}
\caption{ \footnotesize{First row: estimated corrections $\Delta\dot\varpi$ to the  Newton/Einstein perihelion precessions of the inner planets, in $10^{-4}$ arcsec cy$^{-1}$  according to Table 3 of \protect\ctz{Pit05} (Mercury, Earth, Mars). The result  for Venus has been obtained by recently processing radiometric data from Magellan spacecraft \protect \ct{PitMAG}. In square brackets we quote the formal, statistical errors resulting from the least-square estimation process. In the text we used the re-scaled errors.
Second row: nominal general relativistic Lense-Thirring precessions $\dot\varpi_{\rm LT}$, $10^{-4}$ arcsec cy$^{-1}$. Such effects, not included in the dynamical force models of the EPM ephemerides and, thus, present, in principle, in the deterimed $\Delta\dot\varpi$, must be subtracted from the corrections $\Delta\dot\varpi$ in order to single out the anomalous exotic effects induced by neither classical mechanics nor standard general relativity.
%Third row: anomalous perihelion precessions $\dot\varpi_{\dot c/c}$, in $10^{-4}$ arcsec cy$^{-1}$, due to $\dot c/c$ as predicted by %\protect\rfr{dc_dot_o}-\protect\rfr{due} using the values of Table \protect\ref{tavo} for $F(e)$ and \protect\rfr{cdot} for $\dot c/c$.
}}
\centering
\bigskip
\begin{tabular}{ccccc}
\hline\noalign{\smallskip}

%{\begin{tabular}{@{}ccccc@{}} \hrule
%
& Mercury & Venus & Earth & Mars\\
\noalign{\smallskip}\hline\noalign{\smallskip}
$\Delta\dot\varpi$      ($10^{-4}$ \asec)      &  $-36\pm 50[42]$ &  $-4\pm 5[1]$    &  $-2\pm 4[1]$   &  $1\pm 5[1]$ \\
$\dot\varpi_{\rm LT}$   ($10^{-4}$ \asec)      &  $-20$           &  $-3$            &  $-1$           &  $-0.3$ \\
%
%$\dot\varpi_{\dot c/c}$ ($10^{-4}$ \asec)      &  $10.7\pm 2.7$  &  $63.6 \pm 16$  &  $20.0 \pm 5$  &  $4\pm 1$\\
%
\noalign{\smallskip}\hline
\end{tabular}\label{tavo2}
\end{table} %
In order to have the fully non-relativistic, exotic effects, the Lense-Thirring precessions, not modelled in the EPM ephemerides, have to be subtracted from the estimated corrections, i.e. one has to use
\eqi\Delta\dot\varpi^{\ast} = \Delta\dot\varpi-\dot\varpi_{\rm LT}.\lb{Dperi}\eqf
Now, $\Delta\dot\varpi^{\ast}$   can fruitfully be compared to \rfr{dc_dot_o}  to constrain $\dot c/c$ by assuming that they are entirely due to the putative dynamical effects due to the first derivative of  $c$.

By letting $\dot c/c$ be a free parameter, we can constrain it by comparing \rfr{dc_dot_o} and Table \ref{tavo} to the estimated $\Delta\dot\varpi^{\ast}$ quoted in Table \ref{tavo2}. From a weighted mean of the values of $\dot c/c$ obtained with the four inner planets it turns out
\eqi\rp{\dot c}{c} = (0.5\pm 2) \times 10^{-7}\ {\rm yr}^{-1},\eqf
compatible with \rfr{cdot}.
Our result, obtained without considering $\Delta\alpha/\alpha$ and valid for the last century, is very conservative and pessimistic: indeed, we did not use the mere formal, statistical errors in $\Delta\dot\varpi$ and we linearly added the errors $\delta\Delta\dot\varpi$ and $\delta\dot\varpi_{\dot c/c}$ in constructing the total uncertainty in $|\Delta\dot\varpi^{\ast}-\dot\varpi_{\dot c/c}|$.
%Such a test is related to the magnitude of $\dot c/c$ which, in turn, depends on the physical phenomenon which is assumed to yield a time-variation of %$c$ and of its detection through other phenomena, as in the case of $\Delta\alpha/\alpha$;
%conversely, %such a bound is compatible with that can be obtained from the constrain in the variation of $\alpha$ over the last 1.8 Gyr
%\eqi\left|\rp{\dot\alpha}{\alpha}\right|\leq 3\times 10^{-17}\ {\rm yr}^{-1}\eqf
%from an analysis of the Oklo mine data \ct{Pet}.

By suitably combining the perihelia of various pairs of planets it is possible to perform a more stringent  test of the hypothesis that currently $\dot c/c\neq 0$, independently of its origin and magnitude. Indeed, \rfr{dc_dot_o} and \rfr{mega} yield a function of $a$ and $e$ which represents a distinctive signature of the dynamical effects of $\dot c/c$, irrespectively of its size; moreover, it is important to note that $\dot c/c$ enters \rfr{dc_dot_o} as a multiplicative factor. Thus, by taking the ratios $\mathcal{A}$ of \rfr{dc_dot_o} for different pairs of planets  A and B it is possible to  construct  theoretical predictions which are, at the same time, independent of the magnitude of $\dot c/c$ and still retain a pattern characteristic of $\dot c/c$ itself. Thus, $\mathcal{A}$ can be compared to $\Pi=\Delta\dot\varpi^{\ast}_{\rm A}/\Delta\dot\varpi_{\rm B}^{\ast}$  for the same pairs of planets: if $\mathcal{A}\neq \Pi$ within the errors, i.e. if $|\mathcal{A}-\Pi|\neq 0$ within the errors for some of  the pairs considered, we must reject the possibility that $c$ is nowadays varying according to $\dot c/c\neq 0$. The results are in Table \ref{cazzo};
\begin{table}
\caption{ \footnotesize{First column: pair of planets A and B. Second column: observationally determined ratios $\Pi=\Delta\dot\varpi^{\ast}_{\rm A}/\Delta\dot\varpi^{\ast}_{\rm B}$ for A and B. Third column: theoretically predicted ratios $\mathcal{A}=F_{\rm A}a_{\rm B}/F_{\rm B}a_{\rm A}$  for A and B.  Fourth column: $\Gamma = |\Pi- \mathcal{A}|/\delta\Pi$; $\Gamma >1$ means that $\Pi\neq\mathcal{A}$ within the errors. It turns out that the uncertainties in $e$ and \protect \ct{PitSS} $a$ are completely negligible in evaluating the errors in $\Pi-\mathcal{A}$ which are, instead, dominated by $\delta\Pi$.
 }   }
\centering
\bigskip
\begin{tabular}{cccc}
\hline\noalign{\smallskip}

%{\begin{tabular}{@{}cccc@{}} \hrule
%
A B & $\Pi$ & $\mathcal{A}$ & $\Gamma$\\
\noalign{\smallskip}\hline\noalign{\smallskip}
Venus Mars & $ -0.8\pm 6.8 $   & 17.3 & 3\\
Earth Mercury & $  0.06 \pm 0.44 $  & 1.87 & 4\\
Venus Mercury  & $  0.06 \pm 0.50 $ & 5.95 & 12\\
\noalign{\smallskip}\hline
\end{tabular}\label{cazzo}
\end{table} %
the hypothesis $\dot c/c\neq 0$ during about the last century must be rejected at more than $3-\sigma$ level.
Also in this case, our test is conservative because we evaluated the uncertainty in $\Pi$ as
\eqi\delta\Pi\leq |\Pi|
\pare{
\rp{\delta\Delta\dot\varpi^{\ast}_{\rm A}}{\Delta\dot\varpi^{\ast}_{\rm A}} +
\rp{\delta\Delta\dot\varpi^{\ast}_{\rm B}}{\Delta\dot\varpi^{\ast}_{\rm B}}
}.\eqf It must be noted that, since $\Delta\dot\varpi$ are observation-related quantities, it is perfectly meaningful to take their ratios $\Pi$;
the fact that $\delta\Delta\dot\varpi/\Delta\dot\varpi>1$ simply means that $\Delta\dot\varpi$  can still have a non-zero value smaller than $\delta\Delta\dot\varpi$ and that $\Pi$ is compatible with zero within the errors.

It maybe interesting to note that the perihelion precession of \rfr{Mague} by Magueijo \ct{perihel} would survive the test of the ratios of the perihelia.

%Our analysis is equally  valid also for the case of a variation, linear in time, of the Newtonian constant of gravitation $G$; indeed, it would be %sufficient to replace $-2\dot c/c$ with $\dot G/G$ in \rfr{picco}. In principle, one should now take into account also the effect of $\dot G/G$ in the %Newtonian monopole term $\bds A_{\rm N}=-GM{\bds r}/r^3$ of the form
%\eqi\Delta\bds A_{\rm N}=\pare{\rp{\dot G}{G}}(t-t_0)\bds A_{\rm N}.\lb{mono};\eqf however, it would not affect the pericentre of a test particle, as %shown in \ctz{AU}, so that our the main conclusion  concerning the impossibility of a non-zero variation of $G$ over the last century would retain its %validity.

\section{Discussion and conclusions}
In this paper we phenomenologically put on the test the hypothesis that the speed of light $c$, with the meaning of space-time structure constant $c_{\rm ST}$, can vary over timescales of about one century. We analytically worked out the dynamical effects induced by a linear variation in time of $c$ on the motion of a test particle orbiting a spherically symmetric body finding that the longitude of pericentre $\varpi$ undergoes secular precessions; a numerical integration of the equations of motion qualitatively confirmed this result. As expected from simple dimensional considerations, the expression obtained for $\dot\varpi$ is proportional to the product of $\dot c/c$ by $(GM/c^2 a)F(e)$, where $F(e)$ is a specific adimensional function of the eccentricity $e$. We compared such a theoretical prediction to the recently estimated corrections to the standard Newtonian/Einsteinian perihelion precessions for the inner planets of the Solar System, obtained by analyzing the last century of data, finding $\dot c/c = (0.5\pm 2)\times 10^{-7}$ yr$^{-1}$. Moreover, by taking the ratios of the computed anomalous perihelion precessions for different pairs of planets we were able to obtain a prediction independent of $\dot c/c$ itself and still retaining a pattern characteristic of it. The confrontation of such predicted ratios with the ratios of the observationally determined corrections to the usual perihelion precessions ruled out the hypothesis that $\dot c/c\neq 0$ in the last century  at $3-12\sigma$ level. Our result is independent of any measured variations of other fundamental constants which could be related to a variation of $c$ itself (with a different meaning like, e.g., that of electromagnetic constant $c_{\rm EM}$).
%Our analysis is valid also in the case of a variation of the Newtonian constant of gravitation $G$ provided that $-2\dot c/c$ is replaced by $\dot %G/G$ throughout the paper: indeed, in this case one should also take into account the perturbation of the Newtonian monopole in addition to that of %the 1PN term, but it turns out that it does not affect the pericenter.
If and when other teams of astronomers will estimate their own corrections to the standard perihelion precessions it will be possible to fruitfully repeat this tests.
%\section*{Acknowledgments}
%I thank Johannes Kepler for interesting and fruitful discussions.

 \section*{Acknowledgments}
 I thank E.V. Pitjeva for useful information concerning the estimated perihelion precessions.
%-----------------------------------------

\end{document}